\documentclass[aps,pre,twocolumn,nofootinbib,showpacs]{revtex4-1}
\usepackage{latexsym}
\usepackage{amsmath,amsfonts,graphicx}
\usepackage{amsbsy}
\usepackage{hyperref}
\usepackage{mathrsfs}
\usepackage{color}
\usepackage{psfrag}
\usepackage{enumerate}
\usepackage{amsmath,amssymb,calc,amsfonts}
\usepackage{latexsym}
\usepackage[utf8]{inputenc}
\usepackage[percent]{overpic}
\DeclareFontFamily{U}{rsfs}{}         
\DeclareFontShape{U}{rsfs}{m}{n}{<5> rsfs5 <6><7> rsfs7          %
  <8><9><10><10.95><12><14.4><17.28><20.74><24.88> rsfs10}{}     %
\DeclareMathAlphabet{\mathfs}{U}{rsfs}{m}{n}                     %
\definecolor{indiagreen}{rgb}{0.07, 0.53, 0.03}
\def\beq{\begin{eqnarray}}
\def\eeq{\end{eqnarray}}

\def\nn{\nonumber\\}

\def\={\stackrel{\Delta}{=}}

\def\lie{\pounds}

\begin{document}
\title{Massless charged particles: Cosmic censorship, and Third law of Black Hole Mechanics} \author{C. Fairoos} \email{fairoos.c@iitgn.ac.in}\author{
  Avirup Ghosh}\email{avirup.ghosh@iitgn.ac.in} \affiliation{Indian
  Institute of Technology, Gandhinagar, 382355, Gujarat , India.}
\author{ Sudipta Sarkar}
\email{sudiptas@iitgn.ac.in} \affiliation{Indian
  Institute of Technology, Gandhinagar, 382355, Gujarat , India.}
  
   \begin{abstract}
  The formulation of the laws of black hole mechanics assumes the stability of black holes under perturbations in accordance with the "cosmic censorship hypothesis"(CCH). CCH prohibits the formation of a naked singularity by a physical process from a regular black hole solution with an event horizon. Earlier studies show that naked singularities can indeed be formed leading to the violation of CCH if a near-extremal black hole is injected with massive charged particles and the back reaction effects are neglected. We investigate the validity of CCH by considering the infall of charged massless particles as well as a charged null shell. We also discuss the issue of third law of black hole mechanics in the presence of null charged particles by considering various possibilities.
   \end{abstract}

  \maketitle
\section{Introduction}
One of the major milestones of black hole physics was the realization that black holes follow laws that are similar in form to the laws of thermodynamics \cite{Bardeen:1973gs}. The initial derivations of these laws were based on classical general relativity and the association with thermodynamics was merely an analogy.  But the discovery of Hawking radiation from black hole event horizon \cite{Hawking:1974sw} gave a precise meaning to the thermodynamic properties of a black hole as the consequence of quantum field theory in curved space-time.

 The second law of black hole mechanics was then formulated in the following way: In a classical process involving the evolution of a black hole from one stationary state to another, the area of the horizon cannot decrease, provided the stress-energy tensor of the in-falling matter satisfies the null energy condition. This is known as Hawking's area theorem \cite{Hawking:1971vc}. However, underlying this proof were the following crucial assumptions. The null generators of the horizon are assumed to be geodesically complete. In cases where they aren't, one at least needs the spacetime to be strongly asymptotically predictable.
 Also, while discussing the second law, it is always assumed that the perturbations of black hole horizons decay in the future and the black hole attains a new stationary state asymptotically.  But, such an assumption is questionable as there is no general proof of the stability of black objects in general relativity. In a gravitational collapse, trapped surfaces are always formed whenever there is a high concentration of matter in a region. Then, singularity theorems demand that there must be a singularity, at a finite time, to the future of a trapped surface. Also, one may appeal to the weak form of the ``cosmic censorship hypothesis" (CCH) of Penrose so that the singularity is always contained within an event horizon. The stability of black objects under perturbation is then a consequence of CCH, demanding that the perturbations do not destroy the horizon. But the validity of cosmic censorship hypothesis is itself an open problem in general relativity. If the CCH turns out to be not true, the existence of naked singularities will have important consequences for the nature of extreme gravity.  Any attempt to validate the CCH in its precise form amounts to dealing with the global existence of solutions of general relativity. A simpler approach could be to find a counter example to CCH by generating solutions with naked singularities by a physical process from a regular solution.
 \\

One such possibility is to create an overcharged Reissner Nordstrom (RN) spacetime i.e., a solution in which the mass of the charged black hole ($M$) is less than the absolute value of the charge ($Q$). The solution contains a singularity which is not dressed in an event horizon. We do not indulge into the question of whether such a solution can be obtained by a Cauchy evolution of some regular initial data, and instead, we ask a different question. Can such a solution be obtained from a regular charged black hole by throwing in some reasonable matter with sufficient charge? The answer is in the negative if we start with an extremal black hole \cite{Wald} . But, it turns out that if one takes a charged `test` particle such that its charge is greater than its conserved energy and assumes that the mass and the charge of the black hole changes additively once the particle has fallen into the singularity. Then overcharging is possible if the initial solution is near extremal \cite{Hubeny:1998ga}. Similar results also hold if we try to overspin a non-extremal Kerr black hole \cite{Richartz:2008xm, Jacobson:2009kt, Barausse:2010ka, Colleoni:2015ena, Colleoni:2015afa}. It was further shown in \cite{Hubeny:1998ga} that if one considers the electromagnetic back reaction, then there exists only a small window for the choice of charge and energy of the particle provided the initial black hole is very close to extremality. However, a much more careful analysis of both the electromagnetic self force and energy of the particle radiated to infinity showed that overcharging is not possible \cite{Zimmerman:2012zu}. For the Kerr case the dissipative part of the gravitational back reaction effect was considered in \cite{Barausse:2010ka} for a subset of particle orbits that can cause overspinning and it was seen that overspinning may not be averted. In \cite{Colleoni:2015ena, Colleoni:2015afa} both the dissipative and the conservative effects were considered and it was shown that overspinning is averted.

On the other hand, for the case of a dynamical collapse of a thin charged shell into the charged black hole, there is no scope for overcharging. The case of overcharging with test fields have also been considered in \cite{Kommemi:2011wh, Natario:2016bay,Toth:2011ab,Toth:2015cda} and the results indicate that overcharging is not possible. 

The study of a possible counterexample of CCH is extremely important to understand the foundation of general relativity. Therefore, we need to explore the process of overcharging a charged black hole in various situations to check the validity of CCH. This is the main motivation of this work, and we deal with the same problem in the context of `massless charged particles`. Though not found in nature, to the best of our knowledge, there is no argument that rules out the existence of such particles unless the particle is of spin $1$ or larger \cite{Case:1962zz, Weinberg:1980kq, Sudarshan:1981cj}. In fact, the existence of massless charged particles in a quantum theory is linked to the complete solution of the problem of collinear infrared divergences in quantum field theory. A consistent classical dynamics of such massless charged particle is given in \cite{Lechner:2014kua}. Also, there are solutions in general relativity, e.g. the charged Vaidya solution that requires a stream of charged massless particles as the matter content. We want to study if we can overcharge a Reissner Nordstrom black hole using such null charged particles. In the absence of a tuneable parameter, namely the rest mass, the bounds obtained in \cite{Hubeny:1998ga} might be drastically different and could yield results which are different.\\

It seems reasonable to assume that charged null particles in a given solution of the Einstein-Maxwell system should follow a null geodesic. However, it was shown that the particle must interact with the electromagnetic field as well, hence modifying the equation of motion that it follows \cite{Ori}. We will follow this in trying to find the trajectory of the massless charged particle in question. We will see that as a result of the modification of the trajectory, there will be a point on the trajectory where the velocity four-vector vanishes. To the future of this point, the path will be determined by the condition that the trajectory remains causal. We also study the case of a null charged shell collapsing into a non-extremal RN and check if it is possible to overcharge the black hole.

As an outcome of the above assertion on the trajectory of null charged particles, it followed that the charged Vaidya solution should be modified so that the weak energy condition is satisfied. It was shown in \cite{Ori} that a complete charged Vaidya solution should be constructed by gluing an ingoing and an outgoing Vaidya solutions along the hypersurface on which the momentum four vectors of the stream of null charged particle vanishes. But, the charged Vaidya solution along with weak energy condition is a system which follows the third law of black hole mechanics asserting that it is not possible to make it extremal in a finite (advanced) time. The modification of the solution as suggested  \cite{Ori}  demands a careful analysis of the issue of the third law.  The original proof of the third law \cite{Israel} was the in the context of the ingoing Vaidya solution only and this needs modifications in this new setting. The results obtained are rather surprising. We find that the black hole can become extremal in finite time, in this modified setting, in certain special circumstances. However the third law seems to remain intact in spirit.  

\section{Massless charged particle and the overcharging problem}
First, let us consider the problem of overcharging a black hole. We start with the metric of a Reissner Nordstrom (RN)  black hole given by,
\beq
ds^2=-f(r)dt^2+\frac{dr^2}{f(r)}+r^2d\Omega^2,
\eeq
where $f(r)=\left(1-\frac{2M}{r}+\frac{Q^2}{r^2}\right)$. The outer and inner horizons are, $r_{\pm}=M\pm\sqrt{M^2-Q^2}$. The electromagnetic potential is $A=-\frac{Q}{r}dt$. Now consider a massless charged particle in this background geometry. The equation of motion of such a particle follow a modified Lorentz force equation \cite{Ori} viz.
\beq
k^a\nabla_ak^b=qF^b_{~c}k^c. \label{nulltrac}
\eeq
where $F_{ab}=2\partial_{[a}A_{b]}$ is the electromagnetic field strength, $k^a$ is the null four-momentum and $q$ the charge of the particle. As suggested by \cite{Ori}, the particle's motion deviates from the null geodesic of the spacetime and the particle does interact with the background electromagnetic field by a Lorentz force. In fact, this deviation from null geodesic motion is necessary if we want the trajectory to be consistent with the field equation \cite{Ori}. This modification has important implication for a charged Vaidya solution. A charged Vaidya solution suffers from a fundamental difficulty that an observer who moves on a timelike geodesic may measure a negative energy density in a certain region of space-time. If we include the Lorentz force term in the trajectory of the null charge particle, it is possible to show that such regions are removed from the physical space-time due to the existence of a bouncing surface for the charged null matter. \\

We would like to know whether such a charged massless particle moving along the trajectory given by Eq. (\ref{nulltrac}) can overcharge a non-extremal Reissner Nordstrom black hole and leads to a naked singularity. To start, note that since $\partial_t$ is a Killing vector viz. $\lie_{\partial_t}g_{ab}=0, \lie_{\partial_t}A_a=0$, it follows that,
\beq
-E=(k_a+qA_a)\partial_t^a
\eeq
is a constant along the the integral curves of $k^a$ which implies that $k^t=\left(E-\frac{qQ}{r}\right) / f(r)$. Using the fact that $k^a k_a=0$ and writing $k^a=\frac{dx^a}{d\lambda}$, we have for a charged massless particle,
\beq
0=-f(r){t'}^2+\frac{{r'}^2}{f(r)}
\eeq
which implies ${r'}^2=\left(E-\frac{qQ}{r}\right)^2$ and the `prime' denotes the derivative w.r.t $\lambda$. The difference between this equation with that of the massive charged case is the absence of a term  $m^2\, f(r)$ in the expression for $ {r'}^2$, where $m$ is the mass and this may lead to changes in the results obtained in \cite{Wald, Hubeny:1998ga}. As it is evident that there are two solutions ${r'}=\pm\left(E-\frac{qQ}{r}\right)$. Choosing a particular branch, initially, implies that the trajectory follows that branch until the critical point where $ r'=0$ is reached. At this point, it is possible to extend the curve in the future and smoothly join to either of the two branches. However, as we will see causality implies that only one branch is preferred. A look at $k^t$ shows that in order for the trajectory to be future-directed, as it crosses the outer horizon $r_+$, it is necessary that,
\beq\label{FDIR}
E>\frac{qQ}{r_+}
\eeq
Now to overcharge a Reissner Nordstrom black hole, one must have: $q + Q  > E + M$ and if we start with an initial extremal solution such that $|Q| = M$, we have a condition that $ q > E$. However for an extremal black hole with $r_{+} = Q$, eq.(\ref{FDIR}) implies $q<E$. Therefore there is a contradiction. Hence it is not possible to overcharge an initially extremal RN black hole by throwing in some null charged particle. This extends the result of \cite{Wald} for the massless case.

Next, consider the case of a non extremal charged black hole, for which the condition for overcharging is $q+Q > M+E$. The critical point is $r_c= q Q / E$. In regions $r < r_c$, $E- (qQ / r)$ is negative. If $r_c > r_+$, then $k^t=\left(E-\frac{qQ}{r}\right) / f(r)$ becomes negative in regions $r_+ < r < r_c$ where $f(r)$ is positive. Therefore, the trajectory becomes past directed before it could reach the horizon. Hence, it is necessary that the critical point lies inside the outer horizon i.e. $r_c < r_+$. Moreover one needs to consider the ingoing branch $ r' =-\left(E-\frac{qQ}{r}\right)$, so that the particle is in-falling.  From the overcharging condition it follows that $ E < q$ which also implies that $r_c > Q$.  Since for non extremal charged black hole, $r_{-} < Q$, we estimate the critical point as $r_c > Q > r_{-}$.
\begin{figure}[h]
\includegraphics[width=\linewidth]{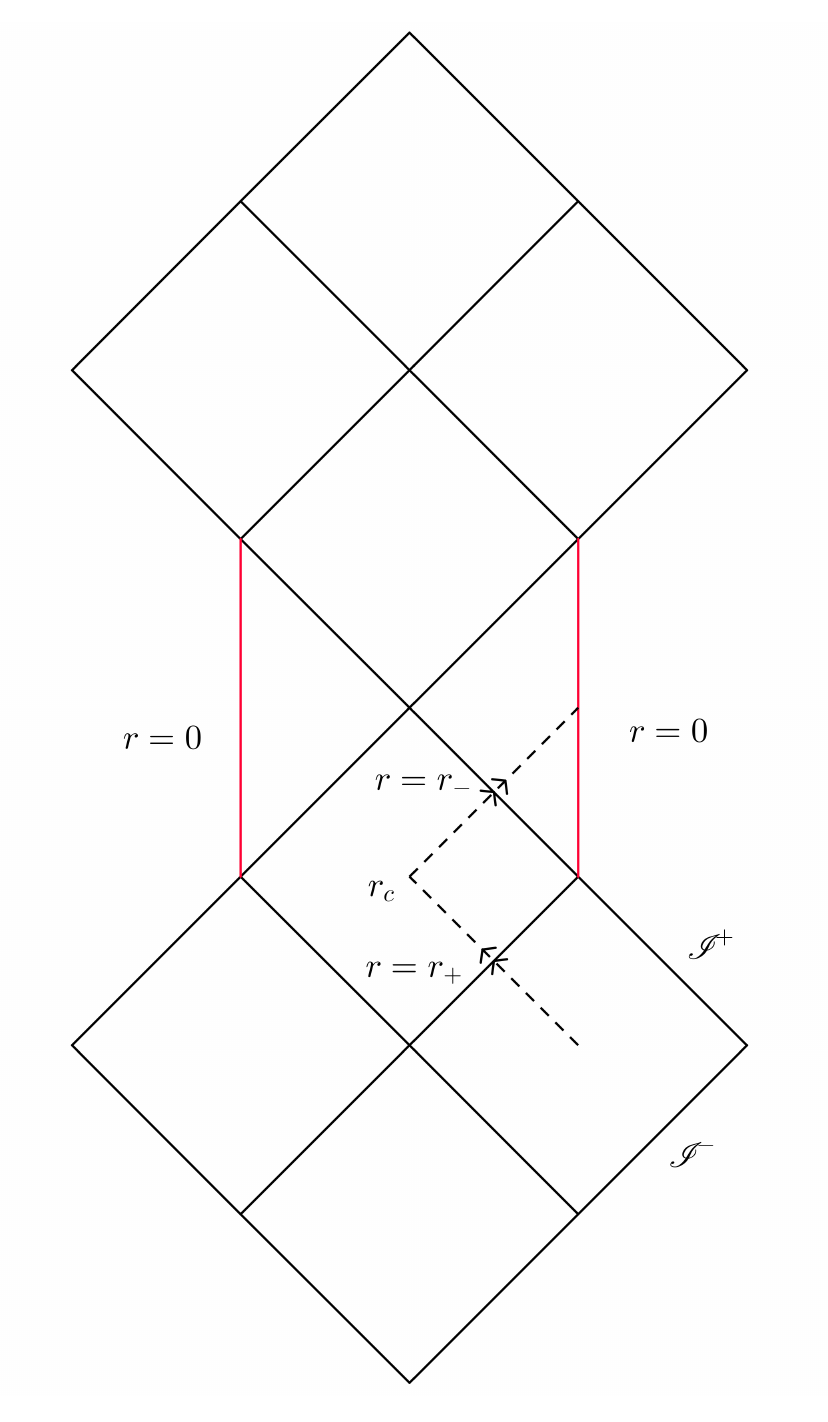}
\caption{The dotted line represents trajectory of the massless charged particle when it's charge (q) and energy (E) satisfies the conditions $q-E>M-Q$ and $E>\frac{qQ}{r_+}$.}
\label{fig.NP}
\centering
\end{figure}
Once the particle has crossed the outer horizon $r=r_+$ the radial coordinate behaves as time. At points $r<r_c$ the factor $E-\frac{qQ}{r}$ becomes negative. Since $r$ is decreasing to the future, one therefore, has  to smoothly join the initial curve for $r>r_c$ with branch $r' =E-\frac{qQ}{r}$ across the point $r=r_c$ so that the curve remains future directed and in-falling. The metric component $f(r)$ is negative in the region $r_-<r<r_+$, so $ t'$ becomes positive beyond $r=r_c$ and the particle therefore follows the dotted curve shown in the figure \ref{fig.NP}. Note that at the point $r=r_c$, $k^a$ is zero and therefore the curve is not `discontinues' as it might appear. It is therefore evident there exists a choice of $q$ and $E$ such that extremization is possible. The choice can be made in the following way. For a given non-extremal RN black hole, choose the charge (q) and then choose $E$ such that $E>\frac{qQ}{r_+}$ and $E<q$. It should be noted that given a non-extremal black hole there always exists such a choice. This situation is however expected to change if back reaction effects are taken into account which we will briefly discuss later. \\
It is important to note that if we do not consider the modification of the trajectory suggested in \cite{Ori}, the massless charged particle would follow a null geodesic and remain future directed always once the energy $E$ is chosen to be positive. Then, the condition $E<q$ is enough for overcharging to take place. Therefore, our result reinforces the modification suggested in \cite{Ori}.

\section{Null charged shell}
Let us now consider a null charged shell of energy $E$ (as measured by an inertial observer) and charge $q$ collapsing into a charged black hole with mass $M$ and charge $Q$. If the shell is spherically symmetric and non-radiating then the spacetime outside the shell can be taken to be an RN with mass $M+E$ and charge $Q+q$. Let us denote the spacetimes inside and outside the shell to be $(\mathcal M_\pm,g_\pm)$ with metrics,
\beq
ds^2=-f_\pm(r)dv^2+2dvdr+r^2d\Omega^2
\eeq
Let the shell be parametrized by coordinates $\lambda,\theta$ and $\phi$. The embedding of the shell in the given spacetimes will then be given by $v=V^\pm(\lambda), r=R^\pm(\lambda)$ where $\lambda$ is a parameter along the shell. The four velocity of the shell will then be given by $U^a=(V^\prime,R^\prime,0,0)$ where the prime represents derivative w.r.t $\lambda$ . For a radially moving future directed null shell, we must have $V^\prime=0$. Therefore the null normal is $l^{\pm a}=(0,R^{\pm\prime},0,0)$ and the transverse null one form is $n^{\pm}_a=(\frac{f^\pm}{2R^{\pm\prime}},-\frac{1}{R^{\pm\prime}},0,0)$. The continuity of the metric across the shell then requires $R^+=R^-$. Though one does not have a unique extrinsic curvature on a null surface, let us take one representative from the class of extrinsic curvatures. Since the normal bundle is one dimensional the components of the extrinsic curvature along $l^a$ is $K_{AB}=g_{a b}l^{a}(\nabla_{\partial_A}\partial_B)^b$, where $A,B$ denotes tangential coordinates. It follows that the non-zero components are,
\beq
K^\pm_{\theta\theta}=-R^{\pm},~~~~~~~~~~K^\pm_{\phi\phi}=-R^{\pm}\sin^2{\theta}
\eeq
As is the case with null shells, it is continuous across the surface  \cite{BH:03,Poisson}. Hence we consider the transverse extrinsic curvatures given by, $\tilde K_{AB}=g_{ab}n^{a}(\nabla_{\partial_A}\partial_B)^b$
\begin{gather}
\tilde K^\pm_{\theta\theta}=\frac{f_\pm \, R^{\pm}}{2R^{\pm\prime}},~~~~~~~~~~\tilde K^\pm_{\phi\phi}=\frac{f_\pm \,R^{\pm}\sin^2{\theta}}{2R^{\pm\prime}}\nn
\tilde K^\pm_{\lambda\lambda}=-R^{\pm\prime\prime}
\end{gather}
The surface stress energy tensor is then given by \cite{BH:03,Poisson},
\beq
t^{ab}=\mu l^a l^b+Pq^{ab},
\eeq
where $q^{ab}$ is the intrinsic metric of the space like cross section of the shell. In our case, we will have $\mu=\frac{f_+-f_-}{RR^\prime}$ and $P=0$. 
The energy as measured by a stationary observer ${\cal E}$ is obtained by contracting the stress energy tensor by the vector $\partial_v$ and this gives,

\beq
 {\cal E}(r)=\frac{R^\prime }{R}\left(\frac{2qQ+q^2}{R^2}-\frac{2E}{R}\right)
\eeq
For $R^\prime$ to be negative for a in-falling shell and the stress energy tensor to satisfy the weak energy condition at $R=r_+$, we need,
\beq\label{cond1}
2E\geq\frac{2qQ+q^2}{M+\sqrt{M^2-Q^2}}\nn
\implies2qQ+q^2-2EM\leq2E\sqrt{M^2-Q^2}
\eeq
On the other hand the overcharging condition $ Q + q > M + E $ implies
\beq\label{cond2}
 2qQ+q^2-2EM>E^2+M^2-Q^2,
\eeq
However, $(\sqrt{M^2-Q^2}-E)^2\geq0$, which means $M^2-Q^2+E^2\geq2E\sqrt{M^2-Q^2}$ and the two conditions, eqn.(\ref{cond1}) and eqn.(\ref{cond2}) are in contradiction with each other. Therefore, it is not possible to overcharge the black hole with a null charged shell as long as the stress tensor of the shell obeys weak energy condition. Note that, in contrast to \cite{Hubeny:1998ga}, where the equations of motion of a time like shell were used, we find that using the weak energy condition is enough to arrive at the contradiction.\\

This ends our analysis of the overcharging problem with a massless charged particle or null shell. In the next section, we consider a charged Vaidya solution and study the validity of the third law of black hole mechanics.

\section{Null charged fluid and the glued Vaidya solution}

Consider the ingoing Vaidya solution,
\beq
ds^2=-f(r)dv^2+2dvdr+r^2d\Omega^2,
\eeq
where $f(r)=1-\frac{2m(v)}{r}+\frac{q^2(v)}{r^2}$. This is a solution of the Einstein Maxwell field equations with the following stress energy tensor.
\beq
T^{ab}=M^{ab}+E^{ab},
\eeq
where $M^{ab}$ is the matter stress energy tensor and $E^{ab}$ is the electromagnetic part. The expression for $E^{ab}$ is standard while the expression for $M^{ab}$ is given by,
\beq
M^{ab}=\frac{1}{4\pi r^2}\left(\dot m-\frac{q\dot q}{r}\right)\delta^a~_r\delta^b~_r=\rho~k^ak^b,
\eeq
where dot denotes derivative w.r.t $v$. If the fluid four velocity is taken to be $k^a = \delta^a~_r $ then it implies that the null charged particles are following an affinely parametrized geodesic and then the weak energy condition is violated at regions $r<\frac{q\dot q}{\dot m}=r_c$ \cite{Ori}. In fact, a time like observer can enter into this region and in principle can measure the local violation of the weak energy condition. This is indeed a problematic feature of the ingoing Vaidya solution. To avoid this pathological nature it was asserted that the null charged particles constituting the fluid must follow a modified Lorentz force equation given by \cite{Ori},
\beq
k^a\nabla_ak^b=qF^b_{~c}k^c,
\eeq
where $q$ is the ratio of the charge density and the energy density. This unavoidably implies that the fluid must become outgoing from being initially ingoing at a critical surface $r=r_c$ hence avoiding the pathological region where violation of the energy condition occurs. As a consequence, one has to glue an ingoing Vaidya to an outgoing Vaidya along the surface $r=r_c$ to recover the full physical space-time. In the foregoing discussions, we will assume that the surface $r=r_c$ is space like. One reason for this consideration is the fact that if the surface is time like, then, both the ingoing and the outgoing Vaidya solutions must coexist in the same region of spacetime and one fails to find the metric that describes this behaviour. This feature also generalized to higher dimensions, as well as with AdS boundary condition and also for higher curvature theories of gravity \cite{Chatterjee:2015cyv}.
\begin{figure}[h]
\center
\includegraphics[width=\linewidth]{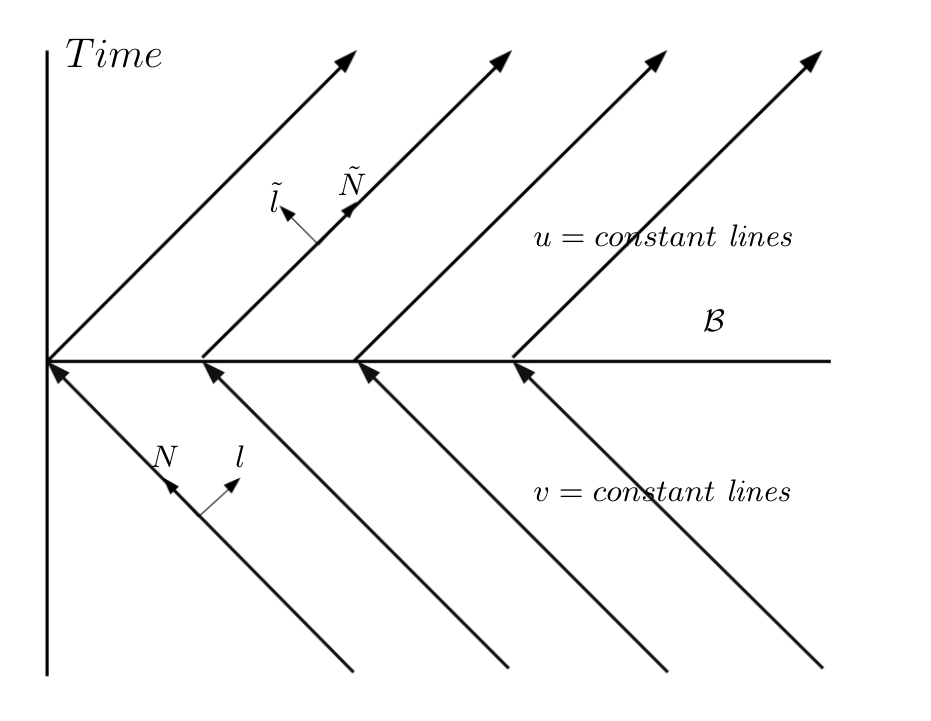}
\caption{The gluing of an ingoing Vaidya and an outgoing Vaidya along a space like surface}
\label{fig.ori}
\end{figure}

Consider two manifolds $\mathcal M,\tilde {\mathcal M}$ corresponding to ingoing and outgoing Vaidya respectively glued along the space like surface $r=r_c=\frac{q\dot q}{\dot m}$, cf. Figure \ref{fig.ori}. The corresponding metrics are given by,
\beq
ds^2=-f(r)dv^2+2dvdr+r^2d\Omega^2\\
ds^2=-\tilde f(r)du^2-2dudr+r^2d\Omega^2.
\eeq
Let the parametric equation for the hypersurface $\mathcal B$, along which the two solutions will be matched be,

\beq\label{parametrization}
r=R(\lambda),~~~~~~~~~~~~~~~v=V(\lambda)\\
r=\tilde R(\lambda),~~~~~~~~~~~~~~~u=U(\lambda)
\eeq
The continuity of the metric then implies that \footnote{The derivatives w.r.t. $\lambda$ are denoted by $^\prime$.},
\beq\label{continuity}
(-f(R) V^{\prime 2}+2 R^{\prime} V^{\prime})d\lambda^2+R^2d\Omega^2\nn
\stackrel{\mathcal B}{=}
(-\tilde f(\tilde R) U^{\prime2}-2{\tilde R}^{\prime}U^{\prime})d\lambda^2+\tilde R^2d\Omega^2,
\eeq
which implies $R(\lambda)=\tilde R(\lambda)$, Further, we must have $f(R)=\tilde f(R)$. This puts the conditions $m(V)=\tilde m(U)$ and $q(V)=\tilde q(U)$ and,
\beq
(-f(R)V^{\prime2}+2R^{\prime}V^{\prime})=(-f(R)U^{\prime2}-2R^{\prime}U^{\prime})
\eeq
Hence, we have the following matching condition,
\beq
-f(R)(V^{\prime}+U^{\prime})(V^{\prime}-U^{\prime})+2R^{\prime}(V^{\prime}+U^{\prime})=0
\eeq
Therefore, one can either have $(V^{\prime}+U^{\prime})$ or $-f(R)(V^{\prime}-U^{\prime})+2R^{\prime}=0$. Following the nomenclature in \cite{Creelman:2016laj} we will call the first matching condition the``Reflective matching" and the second one the ``Ori matching". In the latter case, if at any time $v=v_0$ and $\lambda=\lambda_0$, an apparent horizon (given by $f(R)$=0) coincides with $\mathcal B$, then it would imply $R^{\prime}\arrowvert_{\lambda_0}=0$, which will subsequently imply that the surface $\mathcal B$ is null. This contradicts our initial assumption that the hypersurface is space like. Since while discussing the issue of third law, the apparent horizons will be allowed to cross $\mathcal B$, we will not be considering the Ori matching condition. There are no such issues if one considers the reflective matching. Also, either of the matching conditions along with the fact that the hypersurface $\mathcal B$ is space like restricts the choice of the functions $m(v)$ and $q(v)$. The existence of such a choice was shown in \cite{Creelman:2016laj} where the authors constructed such functions from the conditions. Here, instead of restricting ourselves to a particular choice of these functions, we address the question, whether the apparent horizons given by $r_\pm=m(v)\pm\sqrt{m(v)^2-q(v)^2}$ can cross $\mathcal B$ for some generic $m(v)$ and $q(v)$. In other words, we want to find the location of the surface $\mathcal B$ w.r.t the two apparent horizons of the two Vaidya spacetimes, and the subsequent outward evolution. These would help us to get hints about the location of the extremal apparent horizon if it at all forms in a finite time. This is discussed in section \ref{LOAH} and in the next section we deal with the third law in this context, using the results derived in section \ref{LOAH}.

\section{Third Law}
The third law of black hole mechanics as formulated in \cite{Bardeen:1973gs} says that the state of zero surface gravity cannot be attained in a finite number of steps. In order to have a sensible notion of the term ``finite no. of steps", one must consider the evolution of a black hole, which necessarily goes through a dynamical stage. But during the dynamical stage, there is no definition of surface gravity. If one alternatively takes Planck's version of the third law of thermodynamics, it would imply that the entropy of the extremal black hole must be zero, which of course is in conflict with the fact that extremal black holes do have a non-zero area and entropy. Hence, the process of extremization and the conditions on geometric quantities correctly describing the notion of an extremal horizon must be formulated carefully in a coordinate-invariant way.

In \cite{Israel:1986gqz}, it was argued that an initially charged black hole would become extremal when the trapped surfaces between the inner and the outer horizon have been squeezed out. In other words, on a given time slice, one is left with a marginally trapped surface with trapped surfaces neither on the inside nor on the outside. To model this situation, let us first construct a local null tetrad $l,n,m, \bar m$ where $l$ and $n$ denote the outgoing and the ingoing null directions respectively. The outer black horizon is characterized by the condition $\theta_l=0$. Since there are no trapped surfaces on the inside of the outer black hole horizon, the expansion $\theta_l$ must be negative inside. Therefore, we must have $\lie_n\theta_l < 0$ on the outer horizon. Similarly, there are no trapped surfaces on the outside of the inner black hole horizon, which imposes the condition $\lie_n\theta_l>0$ on the inner horizon. For the extremal black hole, there are no trapped surfaces either on the inside or on the outside of the horizon and therefore the derivative of the expansion must be zero. Hence one defines an extremal black hole horizon by the conditions $\theta_l=0$ and $\lie_n \theta_l=0$. For a spherically symmetric charged Vaidya solution, this condition is equivalent to $m(v)=\lvert q(v)\lvert$ \cite{Booth:2015kxa}.\\

Now, recall the original derivation of the third law as in \cite{Israel}. The proof is by contradiction. The initial assumption can be restated as the black hole extremizes in a finite time $v=v_0$ and was non-extremal for $v<v_0$. The proof then goes by showing that if the weak energy condition were true when $r>r_c$ then the black hole must have been overcharged for $v>v_0$. Note that in this geometry the surface $r=r_c$ lies at the boundary of the region, where the energy condition holds and therefore it is not necessary to consider the case, where the black hole might extremize at $r=r_c$. In the modified solution of \cite{Ori}, the surface $r=r_c$ lies at the interface of the ingoing and outgoing regions and one must, therefore, consider this case as well.

Our first aim here would be to track the outer evolutions of the two apparent horizons of both the ingoing and the outgoing Vaidya solutions in the glued spacetime. To achieve this we will start with the following initial configurations or locations of the apparent horizons and check if they can evolve across the hypersurface $\mathcal B$. We enumerate the cases for the ingoing solution here. Those for the outgoing solution are similar. Let us denote the outer and inner apparent horizons for the ingoing Vaidya solution by $r_+$ and $r_-$ respectively.

\subsection{Location and evolution of the apparent horizons}\label{LOAH}
Note that the location of $\mathcal B$, say at $v=v_0$ can be determined by calculating the value of the metric function $f(R(\lambda))$. For example, if $f(R(\lambda))$ is positive then either $R(\lambda)>r_+>r_-$ or $R(\lambda)<r_-<r_+$. Its location at $v>v_0$ can then be determined by finding the derivative of $f(R(\lambda))$ along $\mathcal B$. We deal with the ingoing part of the solution only. The outgoing piece gives similar results, with appropriate modifications.

\begin{figure}[h]
\includegraphics[width=\linewidth]{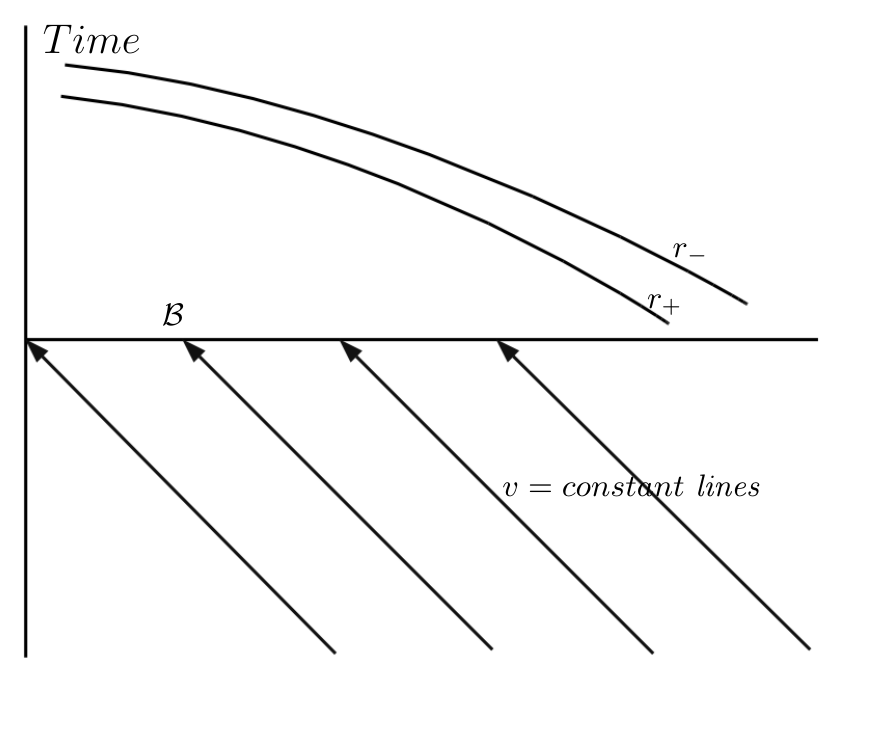}
\caption{Schematic diagram corresponding to $Case-Ia$. The results of section \ref{LOAH} show that the apparent horizons can not cross $ \mathcal{B}$.}
\label{fig.case1a}
\centering
\end{figure}

Consider the ingoing solution. The normal one form to $\mathcal B$ in $\mathcal M_+$ is then given by $dr-\dot r_cdv$. The tangent space of $\mathcal B$ is therefore spanned by $e_1=\dot r_c\partial_r+\partial_v,e_2=\partial_\theta,e_3= \partial_\phi$. If one defines intrinsic coordinates as in eqn \ref{parametrization}, then the push forward of the vector field $\partial_\lambda$ tangent to $\mathcal B$ is $\dot V\partial_v+\dot R\partial_r$. This implies,

\beq\label{fderivative}
\frac{\partial f}{\partial\lambda}\stackrel{\mathcal B}{=}V^{\prime}\partial_{v}f+R^{\prime}\partial_rf=\frac{R^{\prime}}{R^3}(2m(V)R-2[q(V)]^2),\nn
\eeq

where $\partial_vf$ has been dropped because it can be shown to be zero on $\mathcal B$. Let's choose $\lambda$ to be such that it increases outwards, i.e., in the direction of spacelike infinity. This convention is equivalent to choosing $V^{\prime}$ to be positive. We will now discuss the evolutions of the apparent horizons with respect to $ \mathcal{B}$,  by considering two different configurations.  We will consider other possibilities later.\\
\\
$\bf{Case-Ia}$: Both the apparent horizons of the ingoing solution are located to the future of $\mathcal B$ at some value of the parameter $\lambda=\lambda_i$, i.e $R(\lambda_i)>r_+(\lambda_i)>r_-(\lambda_i)$, cf. Figure \ref{fig.case1a}. In this situation, we say that both $r_+$ and $r_-$ are in the unphysical region at $\lambda=\lambda_i$. This nomenclature is motivated by the fact that the spacetime to the future of $\mathcal B$ is the outgoing Vaidya rather than being ingoing Vaidya. We then ask the following question. Is it possible that at some $\lambda > \lambda_i$, $r_+(\lambda)> R(\lambda)$ while demanding that $\mathcal B$ continues to be space like? This case is discussed extensively below.\\

Lets consider the ingoing solution. Suppose at $\lambda=\lambda_0$, $R(\lambda)=r_+(V(\lambda))$ such that in a neighbourhood of $\lambda_0$, $R(\lambda)>r_+(V(\lambda_0))$ for $\lambda<\lambda_0$ and $R(\lambda)<r_+(V(\lambda))$ for $\lambda>\lambda_0$ . This implies the following,
\beq
f(R(\lambda))>0~~~~~~~~~ \lambda<\lambda_0\nn
f(R(\lambda))=0~~~~~~~~~ \lambda=\lambda_0\nn
f(R(\lambda))<0~~~~~~~~~ \lambda>\lambda_0,
\eeq
which also implies that $\frac{df}{d\lambda}\arrowvert_{\lambda_0}<0$. From the expression for the derivative of $f(R)$ derived in eq.(\ref{fderivative}) and the fact that $R(\lambda_0)>r_+(\lambda_0)>m(V(\lambda_0))$ one can infer that at $\lambda=\lambda_0$
\beq\label{R'}
\frac{dR}{d\lambda}<0.
\eeq
Since we have chosen $ V^\prime>0$, eq.(\ref{R'}) together with eq.(\ref{continuity}) implies that $\mathcal B$ ceases to be space like, which contradicts our initial assumption.\\

\begin{figure}[h]
\includegraphics[width=\linewidth]{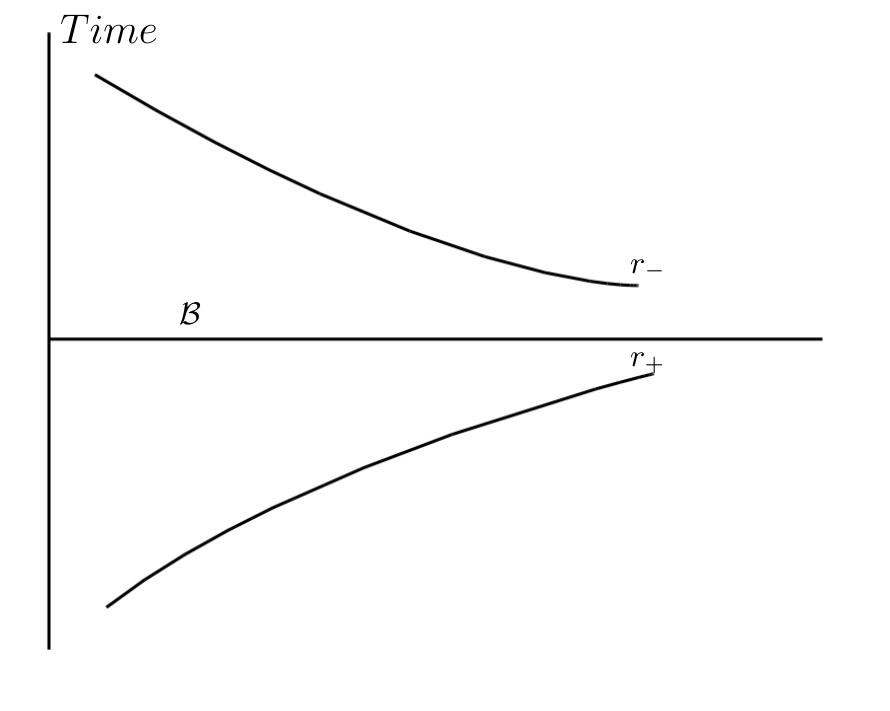}
\caption{Schematic diagram corresponding to $Case-Ib$. The results of section \ref{LOAH} show that apparent horizon $r_-$ can not cross $ \mathcal{B}$ but can however approach $\mathcal B$.}
\label{fig.case1}
\centering
\end{figure}

$\bf{Case-Ib}$: One of the apparent horizons of the ingoing solution is located to the future of $\mathcal B$ at some value of the parameter $\lambda=\lambda_i$, i.e $r_+(\lambda_i)>R(\lambda_i)>r_-(\lambda_i)$, cf. Figure \ref{fig.case1}. In this situation, we say that $r_-$ is in the unphysical region at $\lambda=\lambda_i$. We then try to find out if it is possible that at some $\lambda > \lambda_i$, $r_-(\lambda)> R(\lambda)$ while demanding that $\mathcal B$ continues to be space like?
Suppose, at $\lambda=\lambda_0$, $R(\lambda_)=r_-(V(\lambda))$, such that in a neighbourhood of $\lambda_0$, $r_+(V(\lambda))>R(\lambda)>r_-(V(\lambda))$ for $\lambda<\lambda_0$ and $R(\lambda)<r_-(V(\lambda))$ for $\lambda>\lambda_0$ . This implies the following,
\beq
f(R(\lambda))<0~~~~~~~~~ \lambda<\lambda_0\nn
f(R(\lambda))=0~~~~~~~~~ \lambda=\lambda_0\nn
f(R(\lambda))>0~~~~~~~~~ \lambda>\lambda_0,
\eeq
which also implies that $\frac{df}{d\lambda}\arrowvert_{\lambda_0}>0$.
Therefore at $\lambda=\lambda_0$
\beq
\frac{dR}{d\lambda}<0.
\eeq
Similar arguments as above show that this also contradicts our assumption. The above two situations studied imply that if the apparent horizons are in the unphysical region of the ingoing solution, then they cannot emerge into the physical region during its outward evolution.\\
\\
In fact, one can get a stronger result on the evolution of the apparent horizons only for the case where $R>r_+>r_-$, initially. Let
\beq
x=2mR-2q^2\\
\implies \frac{x+2q^2}{2m} = R>m\\
\implies x>2m^2-2q^2>0
\eeq
As per our convention, $V^{\prime}$ is positive. Since $ f(R)$ is positive in this case, it follows from eq.(\ref{continuity}) that $R^{\prime}$ has to be positive for $\mathcal B$ to remain spacelike. Hence  according to eq.(\ref{fderivative}), $\partial_\lambda f$ is
positive. Therefore $f(R)$ continues to increase, consequently the apparent horizon $r=r_+$ can't even approach $\mathcal B$ in the outward direction. However, note that if $r_+>R>r_-$ initially, then $r_-$ may approach $\mathcal B$. This is because, according to eq. (\ref{continuity}), $R'$ can be chosen to be negative or positive. \\ 

\begin{figure}[h]
\includegraphics[width=\linewidth]{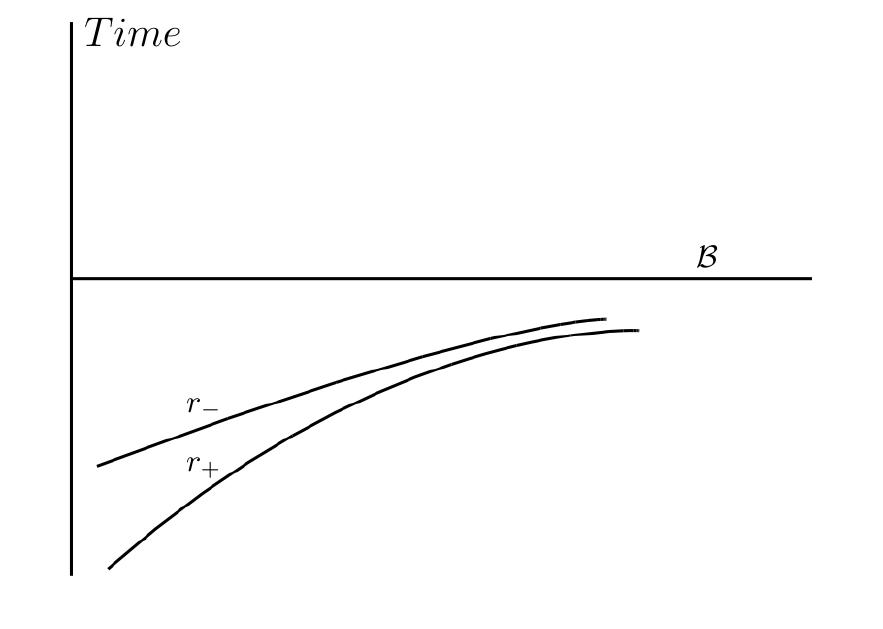}
\caption{Schematic diagram corresponding to $Case-IIa$. It is possible for this evolution to occur.}
\label{fig.case2}
\centering
\end{figure}

$\bf{Case-IIa}$: Both the apparent horizons of the ingoing solution are located to the past of $\mathcal B$ at some value of the parameter $\lambda=\lambda_i$, i.e $R(\lambda_i) < r_-(\lambda_i)<r_+(\lambda_i)$, cf. Figure \ref{fig.case2}. In this case, we will say that both the apparent horizons are located in the physical region of the spacetime. Then the relevant question is whether at some $\lambda>\lambda_i$, $r_-(\lambda)<R(\lambda)$ keeping $\mathcal B$ space like? These questions are addressed below.\\

Suppose at $\lambda=\lambda_0$, $R(\lambda)=r_+(V(\lambda))$ such that in a neighbourhood of $\lambda_0$, $r_-(V(\lambda))<R(\lambda)<r_+(V(\lambda))$ for $\lambda<\lambda_0$ and $R(\lambda)>r_+(V(\lambda_0))$ for $\lambda>\lambda_0$ . This implies the following,
\beq
f(R(\lambda))<0~~~~~~~~~ \lambda<\lambda_0\nn
f(R(\lambda))=0~~~~~~~~~ \lambda=\lambda_0\nn
f(R(\lambda))>0~~~~~~~~~ \lambda>\lambda_0,
\eeq
which also implies that $\frac{df}{d\lambda}\arrowvert_{\lambda_0}>0$.
Therefore at $\lambda=\lambda_0$
\beq
\frac{dR}{d\lambda}>0
\eeq
Following the same arguments as above, we conclude that this is a possible transition.\\

$\bf{Case-IIb}$: One of the apparent horizons of the ingoing solution is located to the past of $\mathcal B$ at some value of the parameter $\lambda=\lambda_i$, i.e $ r_-(\lambda_i)<R(\lambda_i)<r_+(\lambda_i)$. The question we ask here is whether at some $\lambda>\lambda_i$, $r_+(\lambda)<R(\lambda)$, while demanding $\mathcal B$ to remain space like?\\

Now suppose at $\lambda=\lambda_0$, $R(\lambda)=r_-(V(\lambda))$ such that in a neighbourhood of $\lambda_0$, $r_+(V(\lambda))>R(\lambda)>r_-(V(\lambda))$ for $\lambda<\lambda_0$ and $R(\lambda)<r_-(V(\lambda))$ for $\lambda>\lambda_0$ . This implies the following,

\beq
f(R(\lambda))<0~~~~~~~~~ \lambda<\lambda_0\nn
f(R(\lambda))=0~~~~~~~~~ \lambda=\lambda_0\nn
f(R(\lambda))>0~~~~~~~~~ \lambda>\lambda_0,
\eeq
which also implies that $\frac{df}{d\lambda}\arrowvert_{\lambda_0}>0$.
Therefore at $\lambda=\lambda_0$
\beq
\frac{dR}{d\lambda}<0
\eeq
Both the above cases imply that if the apparent horizons are in the physical region of the ingoing solution, then it is possible for them to move into the unphysical region during the outward evolution.

According to the discussion in this section we conclude the following: if any of the apparent horizons are located in the unphysical region, it cannot evolve into the physical region during its outward evolution. The reverse is however possible.

\subsection{Extremization}
 Here,  we explore the process of extremization in the light of the conclusions of the previous section. The cases here will be numbered according to whether the apparent horizons extremize to the future of $\mathcal B$, to the past of $\mathcal B$ or on $\mathcal B$ and has apparently no connection to the numbering in the previous section. However, the results of section \ref{LOAH} will narrow down the number of sub-cases we must deal with. The discussion will only be for the possible cases in the ingoing solution. Those for the outgoing solution are a straightforward modification of those in the ingoing case.\\

Before going over to the arguments, we derive certain relations which will be helpful. Let $\alpha=m^2-q^2$. Then the following equations can be obtained,
\begin{gather}\label{alpha}
\frac{d\alpha}{d\lambda}=2(\dot mm-\dot qq)\frac{dV}{d\lambda}\\
\frac{d^2\alpha}{d\lambda^2}=(2\dot m^2+2m\ddot{m}-2\dot q^2-2q\ddot{q})\left(\frac{dV}{d\lambda}\right)^2\nn
+2(\dot mm-\dot qq)\frac{d^2V}{d\lambda^2}
\end{gather}

Let us now consider each of the possible cases separately.

\subsubsection{{\bf Case 1}}
Suppose, the black hole extremizes to the past of $\mathcal B$. From the considerations of section \ref{LOAH}, it is clear that there is only one configuration, that of Case IIa, whose outward evolution can lead to this. It is only possible if $R(\lambda)<r_-(\lambda)<r_+(\lambda)$ at $\lambda=\lambda_i$ and the black hole extremizes for some $\lambda>\lambda_i$. Note that this situation has been discussed in \cite{Israel}. However, the next two cases are only specific to the kind of solution we are considering, i.e., the glued Vaidya solution. \\

\begin{figure}[h]
\includegraphics[width=\linewidth]{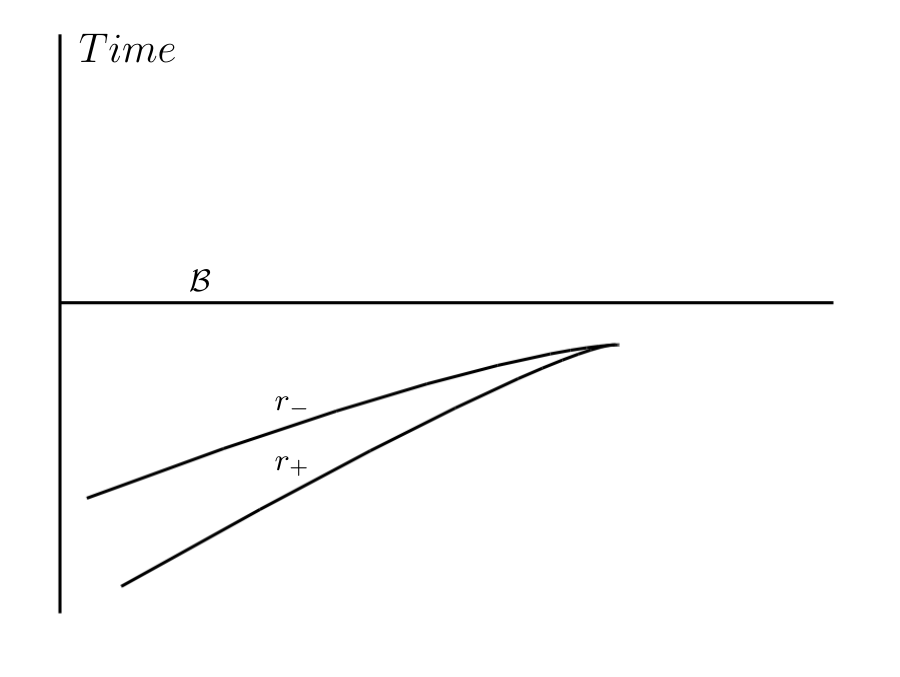}
\caption{Case 1: Black hole extremizes to the past of $\mathcal B$ in a finite time. We have showed that this is not possible.}
\label{fig.c1}
\centering
\end{figure}

Let us assume that the black hole extremizes at $\lambda=\lambda_0$ and $v=V(\lambda_0)=v_0$ and that it was nonextremal i.e $m(v_0)>q(v_0)$ for $\lambda<\lambda_0$. Then $m(v_0)=q(v_0)$ and $r_+(v_0)=m(v_0)$. Suppose $r_+(v_0)>R(\lambda_0)$, cf. Figure \ref{fig.c1}. In that case 
\beq
m(V(\lambda_0))>\frac{q\dot q}{\dot m}\arrowvert_{\lambda_0},\\
\frac{dm(V(\lambda))}{d\lambda}\arrowvert_{\lambda_0}>\frac{d q(V(\lambda))}{d\lambda}\arrowvert_{\lambda_0}.
\eeq
If we choose $\frac{dV}{d\lambda}>0$, so that $\lambda$ increases with $v$, then the above condition  
implies that $m(v)<q(v)$ to the past of $v_0$ which contradicts our assumption. Note that this argument is essentially the same as that given in \cite{Israel}.

\subsubsection{{\bf Case 2}}
The black hole extremizes to the future of $\mathcal B$. In this case, there are several possibilities. However, in either case, the extremal horizon is located in the unphysical region. Therefore, we do not discuss it further. However, we do show that such an evolution, though possible, is unphysical.\\

Suppose the black hole extremizes at $\lambda=\lambda_0$ and $v=V(\lambda_0)=v_0$ and that it was nonextremal before that i.e $m(v)>q(v)$ for $v<v_0$. Then $m(v_0)=q(v_0)$ and $r_+(v_0)=m(v_0)$. Suppose $r_+<R(\lambda)$. In that case 
\beq
m(V(\lambda_0))<\frac{q\dot q}{\dot m}\arrowvert_{\lambda_0},\\
\frac{dm(V(\lambda))}{d\lambda}\arrowvert_{\lambda_0}<\frac{d q(V(\lambda))}{d\lambda}\arrowvert_{\lambda_0}\\
\dot m(v)\arrowvert_{v_0}<\dot q(v)\arrowvert_{v_0},
\eeq 
which implies that $m(v)<q(v)$ to the future of $v_0$ which does not contradict our assumption. However since this piece of the horizon lies in the excised piece of spacetime and is not physical. 

\begin{figure}[h]
\begin{minipage}[h]{\linewidth}
\includegraphics[width=\linewidth]{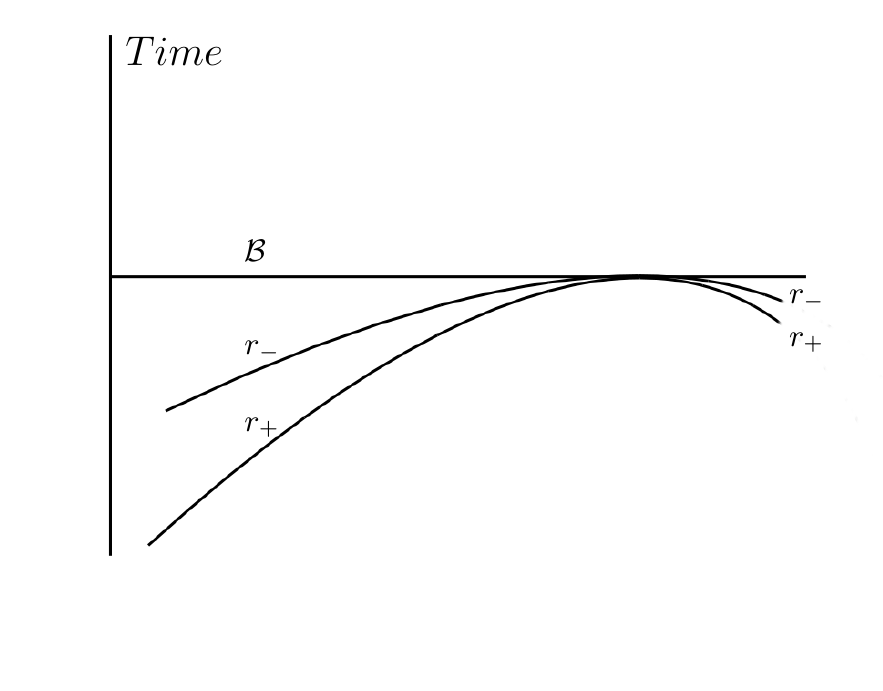}
\caption{Case 3: Black hole extremizes on $\mathcal B$. Apparent horizons bounce back to physical region.}
\label{fig.final1}
\end{minipage}
\hfill
\begin{minipage}[h]{\linewidth}
\includegraphics[width=\linewidth]{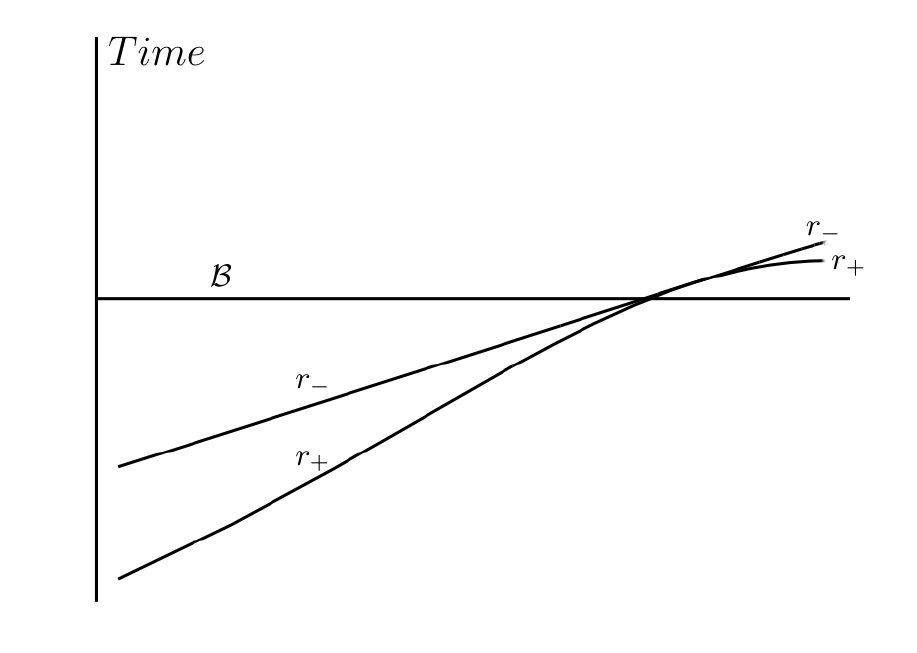}
\caption{Case 3: Black hole extremizes on $\mathcal B$. Apparent horizons continue to evolve into the unphysical region.}
\label{fig.final2}
\end{minipage}
\end{figure}

\begin{figure}
\begin{minipage}[h]{\linewidth}
\includegraphics[width=\linewidth]{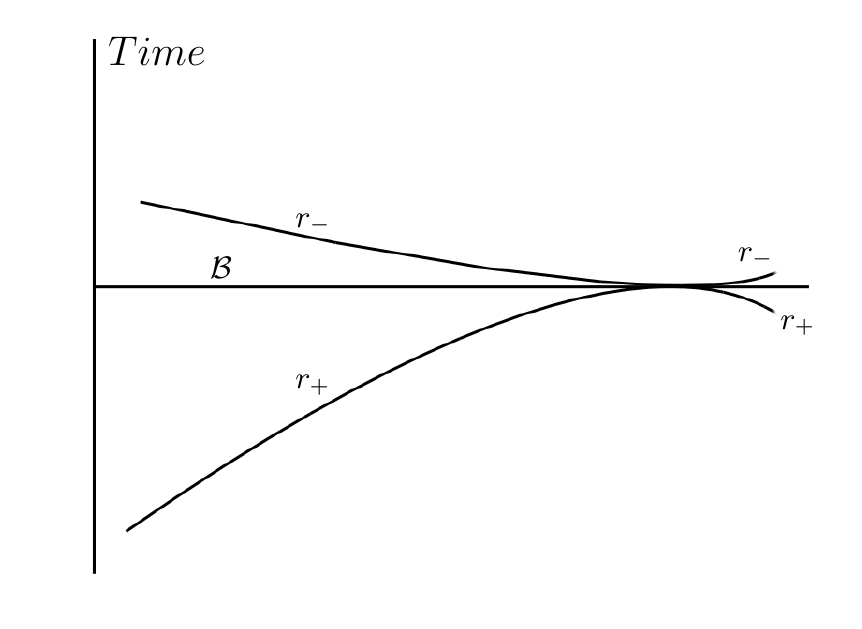}
\caption{Case 3: Black hole extremizes on $\mathcal B$ and  the condition $r_+(\lambda_i)>R(\lambda_i)>r_-(\lambda_i)$ continues to hold.}
\label{fig.ob}
\centering
\end{minipage}

\end{figure}

\subsubsection{{\bf Case 3}}
The black hole extremizes on $\mathcal B$. In this case too, a lot of possibilities are ruled out. The case Ia, for example, is ruled out by section \ref{LOAH}. Hence we are left with two possibilities $r_+(\lambda_i)>R(\lambda_i)>r_-(\lambda_i)$ (Case Ib and IIb) and $r_+(\lambda_i)>r_-(\lambda_i)>R(\lambda_i)$ (IIa), which we need to analyze these in detail.\\

Suppose the black hole extremizes at $\lambda=\lambda_0$ and that $v=V(\lambda_0)=v_0$ and  it was nonextremal before that i.e $m(v)>q(v)$ for $v<v_0$. Then $m(v_0)=\arrowvert q(v_0)\arrowvert$ and $r_+(v_0)=m(v_0)$. Suppose $r_+=R(\lambda)$.

\beq
m(V(\lambda_0))=\frac{q\dot q}{\dot m}\arrowvert_{\lambda_0},
\eeq 
therefore we are unable to assert anything. Now consider the function $\alpha$. By the above conditions,
\beq
\alpha\arrowvert_{v_0}=0\\
\dot \alpha\arrowvert_{v_0}=0\nn
\eeq
Now, suppose the black hole was non extremal for $v<v_0$. Then one has to choose,
\beq\label{alphacondition}
\ddot\alpha\arrowvert_{v_0}\geq 0
\eeq
which implies
\beq
\frac{dR}{d\lambda}=\left(\dot m+\frac{q\ddot q-m\ddot m}{\dot m}\right)\frac{dV}{d\lambda}
\eeq
In this case one can check that $\frac{\partial f}{\partial\lambda}=0$. Therefore one needs to check the sign of $\frac{\partial^2 f}{\partial\lambda^2}$. Note that,
\beq
\frac{\partial^2 f}{\partial\lambda^2}\arrowvert_{v=v_0}=\frac{\dot R}{R^3}\left(\frac{m(q\ddot q-m\ddot m)}{\dot m}\right)\frac{dV}{d\lambda}\arrowvert_{v=v_0}
\eeq
From the expression of $\alpha$ in eq.(\ref{alpha}) and the condition on it in eq.(\ref{alphacondition}) it follows that $\frac{\partial^2f}{\partial\lambda^2}>0$ for $\dot m<0$. In this case $f$ has a minimum at $\lambda=\lambda_0$, which implies $f$ is positive for both $\lambda<\lambda_0$ and $\lambda>\lambda_0$. This means that $\mathcal B$ is either to the future or to the past of both the apparent horizons for  $\lambda<\lambda_0$ and $\lambda>\lambda_0$. The two possibilities, which are consistent with the results of section \ref{LOAH} are schematically portrayed in figure \ref{fig.final1} and \ref{fig.final2}.\\


On the other hand if $\dot m>0$, then $\frac{\partial^2f}{\partial\lambda^2}<0$ which implies $f$ has a maximum at $\lambda=\lambda_0$. Therefore $f$ is negative for both $\lambda<\lambda_0$ and $\lambda>\lambda_0$, and the figure \ref{fig.ob} is implied.

However note that the case where $\dot m<0$, can be removed by simply saying that it involves an influx of negative mass. The other case $\dot m>0$ can be removed by observing that the instantaneous extremal horizon  formed, can never be the outermost marginally trapped surface on $\mathcal B$ and therefore will never be observable by an asymptotic observer. Hence though extremal horizons are formed in finite time it does not seem to violate the third law in spirit.





\section{Conclusions}\label{con}
In the first part of the work, we have considered a charged massless particle which is falling into a black hole and attempted to overcharge it. It turns out that if one starts from an initially extremal one, it is not possible to do so. The black hole does not capture the particles with the energy and charge required to overcharge. However if one initially starts with a non-extremal one, then it is possible to overcharge it. The interpretation seems to be that the original non-extremal black hole jumps to an overcharged one while avoiding the extremal stage. The bounds obtained for the energy and charge of the particle arise only from the overcharging condition and the condition that the particle trajectory remains causal. Further constraints on the allowed choices of energy and charge are expected to occur if one considers the back reaction effects.  We leave the consideration of the back reaction effects for some future work.

We do however consider the case of a null charged shell imploding into the black hole. It turns out that it is not possible to overcharge the black hole with such a charge configuration.

Finally, we conclude that the null charged particles must follow a modified equation of motion as opposed to a geodesic motion, and therefore the issue of the third law of black hole mechanics needs to be readdressed in the context of the charged Vaidya solution as constructed in \cite{Ori} and we note that it is possible to extremize the black hole in a finite time. We discuss further about this in the discussion section.

\section{Discussion}
Establishing the laws of black hole mechanics requires the Cosmic Censorship Hypothesis to be true. It is worthwhile to test the validity of the assumptions behind the `area theorem,' most importantly, the Cosmic Censorship hypothesis. However, it 's hard to either prove or disprove the CCH from the global analysis of the Einstein equations. The alternative approach, therefore, is to look for counterexamples, if any. One such route is to create an overcharged or over spinning black hole from some regular initial solution. These overcharged solutions have naked singularities which are not covered by the horizon. Though these are exact solutions of Einstein's equations, it is not known if they can be obtained through the evolution of some regular initial data. If they could, then they would provide the counterexamples of the weak form of the CCH. In this paper, we, therefore, look into such a scenario for highly boosted or null charged matter falling into a Reissner Nordstrom black hole. There is no evidence of a null charged particle in nature, but our motivation is to check the validity of CCH under extreme conditions and all forms of possible matter. In the absence of a full global analysis of CCH, such studies may provide important clues of the domain of applicability of CCH.
 
We have considered the in-falling matter to be a massless charged particle, treating it as a test particle, we computed its trajectory and the results have been discussed in section \ref{con}. This seems to be a reasonable starting point despite the fact that the notion of point particles is ill-defined both in gravity and electromagnetism. It is because the gravitational or electromagnetic field of a point particle diverges at the location of the particle. One can, however, extract a finite part of this field, and calculate the finite back reaction effects and do modifications to the trajectory. For massive charged particle such effects have been studied and the conclusion is that the CCH holds when the self force effects are taken into consideration \cite{Zimmerman:2012zu}.  In the case of massless particles, it seems that there is neither electromagnetic radiative effects nor conservative self force effects that would modify the trajectory \cite{Lechner:2014kua} and hence only the gravitational back reaction effect caused by the stress energy tensor of the particle can act as a cosmic censor. 

An exact calculation is, however, possible by considering an imploding null charged shell. It should be noted that the `equations of motion' of the time like charged shell were used in \cite{Hubeny:1998ga}. In our case, however, we find that without explicit use of the equations of motion for the charged shell, one can argue that overcharging by a null charged shell is not possible. We establish this only by using the condition that the shell's stress-energy tensor, calculated from the discontinuities in the transverse extrinsic curvatures, must satisfy the weak energy condition.

As has been discussed, the interpretation of overcharging by a particle seems to imply that the extremal state of the black hole is avoided and it jumps from an initially non-extremal state to an overcharged one. Hence it is important to ask if it is indeed possible for an extremal charged apparent horizon to form in a finite time. This is the motivation of the second part of our work, where we investigate the issue of third law of black hole mechanics in the case of the modified charged Vaidya solution as constructed in \cite{Ori}. We found that there is a possibility of an extremal apparent horizon to form momentarily on the hypersurface ($\mathcal B$) where the null charged fluid makes a transition from being ingoing to outgoing. Note that these are consequences of choosing the reflective matching of the coordinates on either side of the surface $\mathcal B$. The configuration should finally evolve into a non-extremal apparent horizon in the excised piece of the solution or bounce back into the physical region, provided $\ddot\alpha$, at extremality, is strictly greater than zero. As has been discussed, inspite of this the third law seems to be unviolated. This is due to the fact that the extremal horizon formed cannot be the outermost  marginally trapped surface on $\mathcal B$. This argument seems to have some familiarity with Israel's other proof of the third law \cite{Israel:1986gqz} in term sof the evolution of the marginally trapped surfaces. Note that, as has been pointed out in \cite{Duztas:2017ycz}, this proof by Isarel, \cite{Israel:1986gqz} already assumes some form of cosmic censorship. \\

One can also look at the consequences if $\ddot\alpha=0$ at extremality and choose $\dddot\alpha<0$. \\

It is known that with the reflective matching the extrinsic curvature has a jump discontinuity as in the case of a thin shell. The effects of this jump on the interpretation of the obtained results might be interesting. It will also important to extend these studies beyond the simple setting of spherical symmetry.\\

While this manuscript was being completed a highly relevant work \cite{Sorce:2017dst} came up, which using the validity of the physical process first law, establishes that a Kerr-Newman Black hole cannot be overspun/overcharged by some generic matter satisfying the null energy condition. Our work related to the null charged particle then is a special case which reinforces their proof. \\

\section{Acknowledgements}
AG is supported by SERB, government of India through the NPDF grant (PDF/2017/000533). SS is supported by the Department of Science and Technology,
Government of India under the SERB Fast Track Scheme for Young
Scientists (YSS/2015/001346). We thank Amitabh Virmani for discussion and suggestions.
The authors also thank the anonymous referees for their valuable comments which has improved the presentation of our results.


\end{document}